\documentclass[pra,10pt,twocolumn,amsmath,amssymb,nofootinbib,bm,bbm,notitlepage,superscriptaddress]{revtex4-1}
\usepackage{tikz} % Package for drawing
\usepackage{amsmath}
\usepackage{hyperref}
\usepackage{braket}
\usepackage{multirow}
\usepackage{xcolor}
\usepackage{dsfont}
\usepackage[thinc]{esdiff}
\usepackage[export]{adjustbox}
\usepackage{hyperref}
\usepackage{physics}
\usepackage{listings}
\usepackage{tikz}
\usetikzlibrary{shapes,positioning}
 \usepackage{float}
\usepackage[ruled,vlined]{algorithm2e}
\usepackage[utf8]{inputenc}
\usepackage{titlesec} 
\usepackage{ulem}

\newcommand{\comment}[1]{}

\newcommand{\im}{\rm{i}} 
\newcommand{\Id}{\mathbb{I}}

\newcommand{\smt}{Science, Mathematics and Technology Cluster, Singapore University of Technology and Design, 8 Somapah Road, 487372 Singapore}
\newcommand{\epd}{Engineering Product Development Pillar, Singapore University of Technology and Design, 8 Somapah Road, 487372 Singapore} 
\newcommand{\ictp}{The Abdus Salam International Centre for Theoretical Physics, Strada Costiera 11, 34151 Trieste, Italy} 
\newcommand{\cqt}{Centre for Quantum Technologies, National University of Singapore 117543, Singapore} 
\newcommand{\majulab}{MajuLab, CNRS-UNS-NUS-NTU International Joint Research Unit, UMI 3654, Singapore} 
\newcommand{\nie}{National Institute of Education, Nanyang Technological University, 1 Nanyang Walk, Singapore 637616}
\newcommand{\quics}{Joint Center for Quantum Information and Computer Science and Joint Quantum Institute, NIST/University of Maryland, College Park, Maryland 20742, USA}

\begin{document}

\title{NISQ algorithm for the matrix elements of a generic observable  } 

\author{Rebecca Erbanni} 
\affiliation{\smt}

\author{Kishor Bharti} 
\affiliation{\quics}

\author{Leong-Chuan Kwek} 
\affiliation{\cqt} 
\affiliation{\nie} 
\affiliation{\majulab} 

\author{Dario Poletti}  
\affiliation{\smt} 
\affiliation{\epd} 
\affiliation{\ictp} 
\affiliation{\majulab}

\begin{abstract}
     The calculation of off-diagonal matrix elements has various applications in fields such as nuclear physics and quantum chemistry. In this paper, we present a noisy intermediate scale quantum algorithm for estimating the diagonal and off-diagonal matrix elements of a generic observable in the energy eigenbasis of a given Hamiltonian. Several numerical simulations indicate that this approach can find many of the matrix elements even when the trial functions are randomly initialized across a wide range of parameter values without, at the same time, the need to prepare the energy eigenstates. %We also study the performance of our algorithm in the presence of noise. 
     %This method does not require the preparation of the energy eigenstates corresponding to which the overlap is to be computed, hence avoiding the associated quantum resource requirements. %This algorithm can be extended to non-normalized anstaz.
\end{abstract}

\maketitle

\section{Introduction}

The landscape of quantum computing technology has shifted dramatically over the previous four decades. Once considered a theoretical endeavour, quantum computing is today a vibrant experimental pursuit. With the advent of noisy intermediate scale quantum (NISQ) devices~\cite{preskill_quantum_2018,bharti2022noisy}, considerable effort has been directed into developing algorithms that can run in the presence of noise on quantum computing systems with 50–70 qubits and restricted qubit connectivity. These algorithms are known as NISQ algorithms, and common examples include the variational quantum eigensolver (VQE)~\cite{peruzzo2014variational,mcclean2016theory,kandala2017hardware} and the quantum approximate optimization algorithm (QAOA)~\cite{farhi2014quantum,farhi2016quantum}. 
% \sout{The recent demonstrations of quantum supremacy by Google and experimentalists at University of Science and Technology of China~\cite{arute2019quantum,zhong2020quantum} have sparked widespread optimism and interest in the potential of NISQ devices, and as a result, significant effort has been directed toward the development of NISQ algorithms for a variety of practical problems.}  

There has been an explosion of recent work on NISQ algorithms to address problems such as finding ground state of Hamiltonians~\cite{peruzzo2014variational,mcclean2016theory,kandala2017hardware,mcclean2017hybrid,kyriienko2020quantum,parrish2019quantum,bespalova2020hamiltonian,huggins2020non,takeshita2020increasing,stair2020multireference,motta2020determining,seki2020quantum,bharti2020quantum,bharti2020iterative,cervera2021meta} quantum simulation~\cite{li2017efficient,yuan2019theory,benedetti2020hardware,bharti2020simulator,barison2021efficient,commeau2020variational,heya2019subspace,cirstoiu2020variational,gibbs2021long,lau2021quantum,Haug_2022,otten2019noise,lim2021fast,lau2021nisq}, combinatorial optimization~~\cite{farhi2014quantum,farhi2016quantum}, quantum metrology~\cite{meyer2020variational,meyer2021fisher,Bharti2021fisher}, and machine learning~~\cite{schuld2019quantum,havlivcek2019supervised, kusumoto2019experimental,farhi2018classification, mitarai2018quantum}. Notwithstanding these research efforts, the practical application of NISQ devices is still a long way off. To attain practical quantum advantage in the NISQ era, it is crucial to examine previously untapped or poorly understood applications. One such application area that may need to receive more attention is calculating off-diagonal matrix elements of generic observables.

Many problems in nuclear physics and other sciences make extensive use of off-diagonal matrix elements. Consider the Brueckner-Bethe-Goldstone equation solution; the nuclear potential is computed from the off-diagonal matrix elements of the Brueckner reaciton matrix~\cite{shang2021exact}. The generalized eigenvalue problems for obtaining energy levels of time-dependent Euclidean correlators~\cite{harris2016variational} in lattice quantum chromodynamics (QCD) rely heavily on information acquired from off-diagonal elements of the matrix of correlators. The rotational-vibrational coupling in quantum chemistry considers the off-diagonal block in the matrix of kinematic coefficients~\cite{gribov2018variational}. Off-diagonal element calculations have also been used to determine the diamagnetic susceptibility and form factor of atoms such as Helium~\cite{shakeshaft1976application}. Given the importance of finding the off-diagonal matrix elements of generic observables, it is pertinent to explore the potential of NISQ  devices for the aforementioned task. Despite the hope that noisy intermediate scale quantum devices will enable the solution of problems in a variety of fields (including quantum chemistry and nuclear physics), the subject of noisy intermediate scale quantum algorithms for off-diagonal matrix element calculation remains poorly understood. One reason for this is that an observable's off-diagonal matrix elements can be a complex number, yet nearly all noisy intermediate scale quantum algorithms are designed to compute real values.

In this paper, we provide a NISQ algorithm for  hybrid classical-quantum computation of the matrix elements (both diagonal and off-diagonal) of a given observable $W$ in the energy eigenbasis.  Indeed, variational principles have been applied successfully in several NISQ algorithms. The Rayleigh-Ritz variational method is used in variational eigensolvers to seek the ground state of a Hamiltonian of a system\cite{peruzzo2014variational}. The McLachlan principle proves useful for dynamical solutions of open quantum systems\cite{cerezo2020variational}. By deriving its motivations from Ref.~\cite{PhysRevA.8.662,RevModPhys.55.725, wadehra1978application}, here we propose a variational principle to find off-diagonal (and diagonal) elements of observables or operators. Our algorithm uses Lagrange multipliers to encode the constraints for the underlying problem into the refined objective. For our model problems, we discuss approaches for both exact and iterative evaluation of Lagrange multipliers. %Our algorithms are applicable to both normalized and non-normalized trial states. 
%We also discuss the performance of our algorithms in the presence of noise. 
%\dpc{noise and non-normalized}
Various numerical simulations suggest that our approach manages to find many of the matrix elements even when one initializes randomly the trial functions over a very broad range of parameters. Given the importance of finding off-diagonal matrix elements for a variety of fundamental problems, we believe that our approaches can be applied to a variety of interesting practical contexts.

We would like to point out that a quantum variational approach for calculating matrix elements was proposed recently in Ref.~\cite{ibe2022calculating}, which relied on the preparation of the energy eigenstates corresponding to which the overlap for the given observable is to be calculated. Unlike this work, our approach does not involve the preparation of the energy eigenstates, hence avoiding the accompanying quantum resource requirements. While our choice of ansatz may seem to resembles existing works in the literature~\cite{mcclean2017hybrid,bharti2020iterative,huggins2020non}, none of these results work for the off-diagonal matrix elements of a generic observable. Moreover, our approach is fundamentally different and uses Lagrange multipliers to encode the constraints for the underlying problem into the
refined objective.

% \kishor{Classical methods to calculate off diagonal matrix elements and their limitations. We can cite Gerjuoy's papers here.}

% \kishor{Using NISQ devices for calculating the off diagonal elements. Vast literature for NISQ algorithms. But no algorithm is known for the off-diagonal matrix elements. Difficulties with calculating the off-diagonal matrix elements}
% \Kwek{Can you have a look at: https://arxiv.org/pdf/2107.02979.pdf:  in this paper they replace off-diagonal elements by diagonal elements in two different bases, $|+\rangle$ and $|y_+\rangle$ but $\langle + | y_+\rangle$ are not orthogonal! Variational principle is then applied to the diagonal elements.}

% \kishor{Contributions of this work}

% In this work, we'll introduce a quantum algorithm for off-diagonal matrix elements for normalized and non-normalized wavefunctions and for Hermitian matrices W that can {\em a priori} be complex. Since the algorithm requires an input matrix that is either real or purely imaginary, one has to add a pre-processing step that maps $W \xrightarrow[]{}W_{re}$ or $W \xrightarrow[]{}W_{im}$ as needed.\\
% \kishor{Highlighting the non-trivialities}

% To clarify the notations, in the following we use $\Id = \left( \begin{array}{cc} 1& 0\\ 0 &1\end{array}\right)$ $X = \left( \begin{array}{cc} 0& 1\\ 1 &0\end{array}\right)$, $Y = \left( \begin{array}{cc} 0& -\im\\ \im & 0\end{array}\right)$, $Z = \left( \begin{array}{cc} 1& 0\\ 0 &-1\end{array}\right)$, and $H_d=\frac{1}{\sqrt{2}}\left( \begin{array}{cc} 1& 1\\ 1 & -1\end{array}\right)$.  

\section{The classical variational algorithms for off-diagonal elements} \label{sec:algo}

Given a system with Hamiltonian $H$ with different eigenenergies $E_i$ and corresponding eigenfunctions $|\phi_i\rangle$ and an observable $W$, what we aim to find are the elements $F_{i,j}=\langle \phi_i |W|\phi_j \rangle$.  % for $i\neq j$ of the observable matrix in the energy eigenfunctions basis.    

One natural approach would be to find the different eigenfunctions of the Hamiltonian and then evaluate the elements, including the off-diagonal ones. However one may not need to do this. In \cite{PhysRevA.8.662,RevModPhys.55.725} the authors showed a variational approach to find such elements which we summarize in the following.  
First we can use a variational ansatz for the eigenfunctions $|\phi_i\rangle \approx|\phi_{i,t} (\vec{\eta}_i)\rangle$ parametrized by the $\vec{\eta}_i$. We can then write a variational function $F_{i,j}^v$ which has zero derivative respect to the $\vec{\eta}_i$ when $F_{i,j}^v=F_{i,j}$. Such approach can readily give both diagonal and off-diagonal elements.  
In the following we consider a normalized parametrization of the trial eigenfunctions $|\phi_{i,t} (\vec{\eta}_i)\rangle$, and an extension to the case of non normalized trial eigenfunctions is discussed in the App.\ref{app:vp_nonnorma}. 

For an observable $W$ which is only real or only imaginary, the variational function $F_{i,j}^v$ is given by 
\begin{align}
    F_v &=\bra{\phi_{i,t}} W\ket{\phi_{j,t}} \nonumber\\  &+\bra{L_{i,a}}(H-E_i)\ket{\phi_{i,t}} +\bra{\phi_{i,t}}(H-E_i)^{\dagger}\ket{L_{i,b}} \nonumber \\ & +\bra{L_{j,a}}(H-E_j)\ket{\phi_{j,t}} +\bra{\phi_{j,t}}(H-E_j)^{\dagger}\ket{L_{j,b}} \nonumber \\ &+\lambda\left[\bra{\phi_{i,t}} W\ket{\phi_{j,t}}\mp \bra{\phi_{j,t}} W\ket{\phi_{i,t}} \right],  
\label{eq:Fijv}
\end{align}
where in the last line one uses the sign $-$ or $+$ depending on whether $W$ is real or imaginary respectively. We note that for a given matrix $W$ we can always write $W_R = (W+W^T)/2$ and $W_I = (W-W^T)/2$, wher $A^T$ is the transposition of $A$, respectively for the real components and for the imaginary ones. In Eq.(\ref{eq:Fijv}) we have used the Lagrange multipliers vectors $|L_{i,\nu}\rangle$ (which, generally, are not normalized) and the scalar $\lambda$. It is thus clear that the terms on the right-hand side of Eq.(\ref{eq:Fijv}) are, apart from the first one, zero for the exact solution, which then results in obtaining the correct evaluation of the off-diagonal term $F_{i,j}$. 

In order to get the expressions for the Lagrange multipliers $L_i$ and $\lambda$, one can expand the first order variation of the functional and set  the coefficients of $\bra{\delta\phi_1}$, $\ket{\delta\phi_1}$,$\bra{\delta\phi_2}$,$\ket{\delta\phi_2}$ to zero. Details of such computations are in App.\ref{app:vp}

We now take a small variation $|\delta\phi_i\rangle$ to the exact eigenfunctions $|\phi_i^{ex}\rangle$, which gives $|\phi_{i,t}\rangle = |\phi_i^{ex}\rangle + |\delta\phi_i\rangle$, and set that the first order corrections to the exact result of the function $F_{i,j}^v$ are zero. This gives expressions for the Lagrange multipliers which are 
\begin{align}
    &\lambda =- 1/2 \\  
    &(H-E_i)|L_{i,\nu}\rangle = -\xi_{R,I}^{\nu} W_{R/I}|\phi_j\rangle/2.   \label{eq:lagr_L}  
\end{align}
with $ \xi_{R,I}^{\nu}=\pm 1$. More precisely, for the real case $\xi_{R}^{\nu}=1$ whether $\nu=a,b$ while for the imaginary case $\xi_{R}^{a}=1$ and $\xi_{R}^{b}=-1$. 
%while it must always be $-W\ket{\phi}$ for both $\ket{L_{1}}$  and $\ket{L_{2}}$, while for the imaginary case, when it's $\ket{L_{1b}}$ there must be - and with $\ket{L_{2b}}$ there must be a +.  
What is important to state, though, is that in principle we do not know the value $E_i$ and thus this will be approximate by the expectation value of the Hamiltonian for that wave function, i.e. $E_i\approx \langle \phi_i| H | \phi_i\rangle$. 

At this point we should solve Eq.(\ref{eq:lagr_L}) for $|L_{i,\nu}\rangle$, but this is not straightforward because $H-\langle \phi_i| H | \phi_i\rangle$ is not invertible. For the purpose of computing these Lagrange multipliers we thus use a modified Hamiltonian 
\begin{align}
    H_{mod,i} = H - \frac{H |\phi_i\rangle \langle \phi_i | H}{\langle \phi_i| H | \phi_i\rangle}
\end{align}
such that the matrix  $H_{mod,i}-\langle \phi_i| H | \phi_i\rangle$ is not singular. 

When it is difficult to evaluate the inverse of $H_{mod,i}-\langle \phi_i| H | \phi_i\rangle$ exactly, it is also possible to implement an iterative approach. As shown in \cite{RevModPhys.55.725}, we can find $|L_{i,\nu}\rangle$ by minimizing 
\begin{align}
M(|L_{i,\nu}\rangle) &= \langle L_{i,\nu} |\left( H_{mod,i} -  \langle \phi_i| H | \phi_i\rangle \right)| L_{i,\nu} \rangle \nonumber \\
&+\langle \phi_j | W_R | L_{i,\nu} \rangle \label{eq:M_for_L_re}   
\end{align}
for the real case, and     
\begin{align}
M(|L_{i,\nu}\rangle) &= \langle L_{i,\nu} |\left( H_{mod,i} -  \langle \phi_i| H | \phi_i\rangle \right)| L_{i,\nu} \rangle
\label{eq:M_for_L_im}   
\end{align}
for the imaginary one. More details can be found in App.\ref{app:vpL}. 

\subsection{Scaling analysis for the overlap calculation}
We consider Hamiltonians $H$ and observables $W$ which can be written as linear combination of poly$(n)$ unitaries where $n$ is the number of qubits over which the Hamiltonian is defined. A typical example is a local spin Hamiltonian which can be written as a sum of polynomially many $n$-qubit Pauli matrices. Furthermore, we consider a  number of ansatz states and Lagrange multipliers that scales polynomially with the system size (number of qubits) as this is often sufficient to obtain accurate results, e.g. using a Krylov basis \cite{bharti2020iterative}.

We thus highlight that there are three types of overlaps that need to be computed for the successful implementation of our NISQ algorithm,
\begin{itemize}
    \item $ \langle \phi_i| H | \phi_i\rangle $: Since the number of terms in the Hamiltonian and the number of ansatz states $\vert \phi_i \rangle$ are polynomially many in number of qubits, the overlaps $ \langle \phi_i| H | \phi_i\rangle $ can be calculated efficiently \cite{PhysRevResearch.1.013006}.
    
    \item $\langle L_{i,\nu} | H_{mod,i} | L_{i,\nu}\rangle$: the estimation of these overlaps requires the calculation of the terms of the form $ \langle \phi_i| H | \phi_i\rangle $, $\langle L_{i,\nu} | H | L_{i,\nu}\rangle$ and $\langle L_{i,\nu} | H | \phi_i \rangle$. Since the number of ansatz states and Lagrange multipliers scale polynomially with the system size, also the aforementioned overlaps can be evaluated efficiently.
    
    \item $\langle \phi_j | W_{R} | L_{i,\nu} \rangle$: Since the operator $W$ can be expressed as a linear combination of polynomially many unitaries, it is easy to see that also the overlaps $\langle \phi_j | W_{R} | L_{i,\nu} \rangle$ can be calculated efficiently. 
\end{itemize}

\section{Results}
We will now show examples which elucidate the effectiveness of an hybrid classical-quantum implementation of this variational approach. We will consider both a single and a two-qubit Hamiltonian and we will use both the exact and the iterative approaches to get the Lagrange multiplier.

\subsection{Models}  

We consider two scenarios. The first scenario is a two level system with Hamiltonian 
\begin{align}
    H_1=X \label{eq:H1}
\end{align}
and $W_1=H_d W^D_1 H_d$ where 
\begin{align}
W^D_1= \begin{pmatrix} 5& 2-2j\\2+2j& 3  \\\end{pmatrix}. \label{eq:Wd1}    
\end{align}
$W^D_1$ is thus the matrix in the energy eigenbasis, as $H_d$ diagonalizes the Hamiltonian $H_1$. 

In the second case we consider the following $2-$ qubit Hamiltonian
\begin{equation}
H_2=2X\otimes \Id + \Id\otimes X + 2Z\otimes X
\label{eq:two_qubits_ham}
\end{equation} 
while we take an $W^D_2$ (therefore in the energy eigenbasis) as the following Hermitian complex matrix 
\begin{equation}
W^D_2=\begin{pmatrix}1&3+1j&5-3j&13+8j\\3-1j& 4&20+5j&25+10j\\5+3j&20-5j&7&6-15j\\13-8j&25-10j&6+15j&10  \\\end{pmatrix}. 
\label{eq:Wd2}
\end{equation} 
The matrix $W_2$ in the computational basis, which we use in the computations, can be obtained from $W^D_2$ and the eigenvalues of $H_2$.

\subsection{Implementation for hybrid classical-quantum computation}    

We test the usefulness of the variational principle from Eq.(\ref{eq:Fijv}) on a hybrid classical-quantum algorithm. Since the solutions can be found where the derivatives are zero, we use a classical optimization algorithm which performs a gradient descent after having evaluated the gradients of $F_{i,j}^v$ over the parameters $\vec{\eta}_i$. The quantum part of the algorithm helps with  the evaluation of the gradients. In practice, one can evaluate all the relevant overlaps once, and then use such knowledge to evaluate the gradients for any given value of the $\vec{\eta}_i$. 
To evaluate the overlaps we write the states $\ket{\phi_{i,t}}$ as 
\begin{equation} 
\label{eq: one_qubit_ansatz}
   \ket{\phi_{i,t}(\theta)}=\cos(\theta_i)\ket{0}+\sin(\theta_i)\ket{1}.   
\end{equation}
for the one-qubit case, and for the two-qubit case we consider the parametrization      
\begin{align}
   &\ket{\phi_{i,t}(\alpha_i,\beta_i,\gamma_i)}=\cos(\alpha_i) \ket{00} +\sin(\alpha_i)\cos(\beta_i) \ket{01}\nonumber \\
   & +\sin(\alpha)\sin(\beta_i)\cos(\gamma_i) \ket{10}+\sin(\alpha_i)\sin(\beta_i)\sin(\gamma_i) \ket{11}. 
\label{eq: two_qubit_ansatz}
\end{align}

Note that here we only consider real trial functions, which is sufficient for our examples, and a generalization to complex ones is straightforward. Furthermore,  we currently use a variational representation of the eigenfunctions which scales linearly with the size of the Hilbert space. In practice, for systems with large Hilbert space, for which a quantum computer would come in handy, one would have to resort to a much smaller parameter space, for example using a limited Krylov basis \cite{lanczos1950iteration,saad1992analysis}.     
For the Lagrange multipliers $\ket{L_{i,\nu}}$ we use an unnormalized ansatz of the form 
\begin{equation} 
\label{eq: L1_ansatz}
   \ket{L_{i,\nu}}=c_{i}\ket{0}+d_{i}\ket{1}    
\end{equation}
for the one qubit case and 
\begin{equation} 
\label{eq: L2_ansatz}
   \ket{L_{i,\nu}}=c_{i}\ket{00}+d_{i}\ket{01} + e_{i}\ket{10} + f_{i}\ket{11}     
\end{equation}
for the qubit case, where the parameters of Eq.(\ref{eq: L1_ansatz},\ref{eq: L2_ansatz}) are real numbers. 

All quantum computations are implemented on the IBM Belem QPU simulator using error mitigation. The overlaps are evaluated each by averaging $50$ estimates of the overlaps each done with $1000$ shots. 

\comment{Consider that, for the purpose of implementing the overlap $\lambda=-\frac{1}{2}\bra{\phi_1}W\ket{\phi_2}$ on a quantum computer, we can write $\ket{\phi_1}=B\ket{00}=(\beta_1 \mathds{1} \otimes\mathds{1}+\beta_2 \mathds{1} \otimes\sigma_x+\beta_3\sigma_x \otimes\mathds{1} +\beta_4 \sigma_x  \otimes\sigma_x)\ket{00}$
 and $\ket{\phi_2}=A\ket{00}=(\alpha_1 \mathds{1} \otimes\mathds{1}+\alpha_2 \mathds{1} \otimes\sigma_x+\alpha_3\sigma_x \otimes\mathds{1} +\alpha_4 \sigma_x  \otimes\sigma_x)\ket{00}$.}

%%%%%%%%%%%%%%%%%%%%%%%%%%%%%%%%%%%%%%%
%%%%%%%%%%%%%%%%%%%%%%%%%%%%%%%%%%%%%%%
\subsection{Single qubit using an exact evaluation of the Lagrange multipliers}  

\begin{figure}[htp]
\includegraphics[width=\columnwidth]{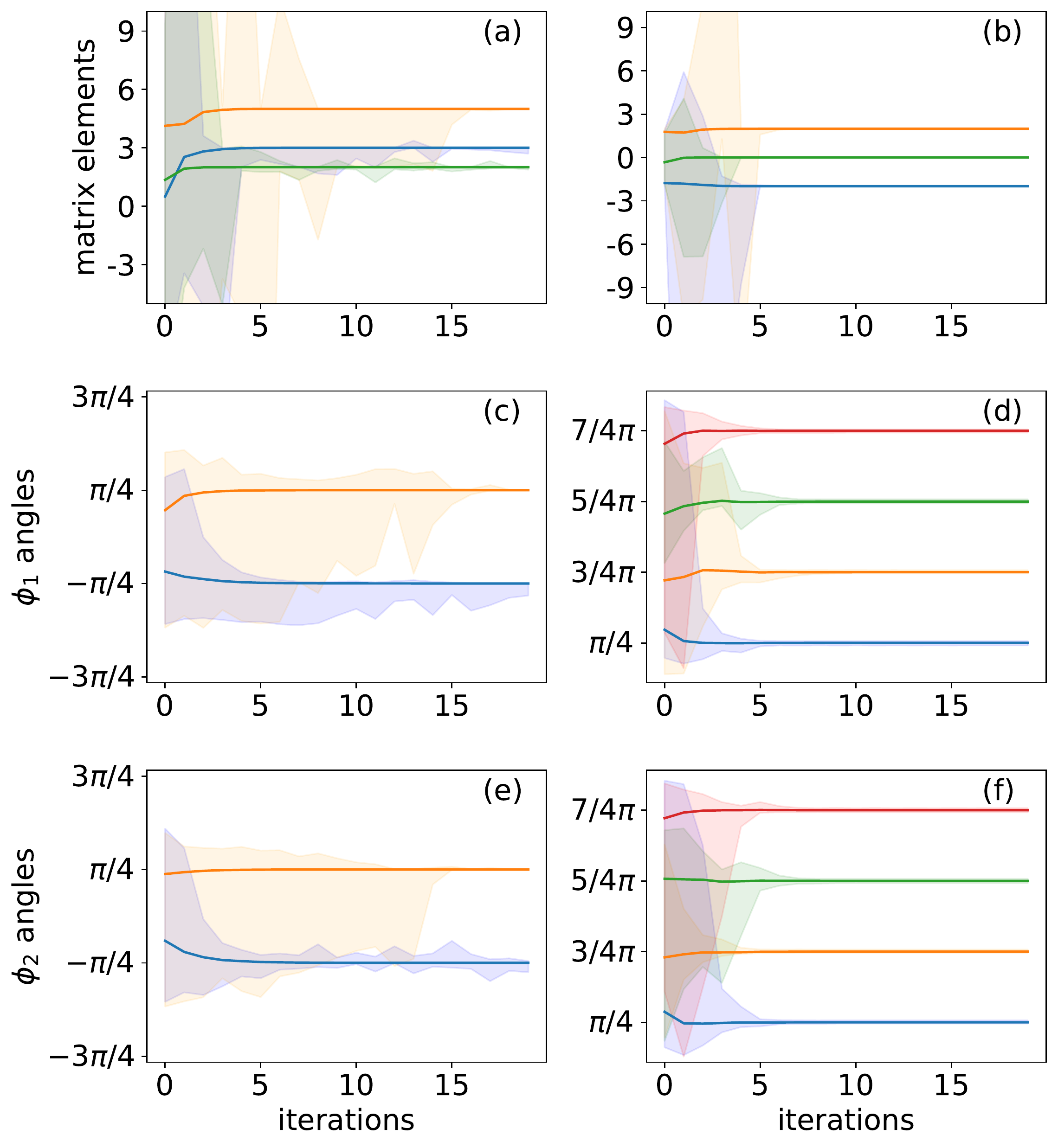}
\caption{Plots for the real and imaginary parts of the matrix elements versus iterations of the classical optimization algorithm, through exact calculation of $\ket{L_{1,\nu}}$ and $\ket{L_{2,\nu}}$ in the 1-qubit case and with error mitigation. Panels (a,c,e) are the results For the real part of $W_1$ in the energy eigenbasis, while (b,d,f) for the imaginary part. Panels (a,b) show the matrix elements while panels (c) to (f) show the angles $\alpha_i$ of the trial eigenfunctions.  
Each panel shows the results from a total of $150$ runs, where the angles have been randomly initialized between 0 and $2\pi$. In all panels, the lines represent the medians of the runs converging to a specific value and the error intervals include $92\% $ of the corresponding runs.} 
\label{fig:method1}
\end{figure}

We now consider the model with a single qubit, with Hamiltonian $H_1$ from Eq.(\ref{eq:H1}) and the matrix $W_1$ derived from the matrix in the energy eigenbasis Eq.(\ref{eq:Wd1}). We use a classical optimization algorithm by deriving analytically the derivatives of Eq.~(\ref{eq:Fijv}) with respect to the angles of Eqs.~(\ref{eq: one_qubit_ansatz}) and the parameters of Eq.~(\ref{eq: L1_ansatz}); the expectation values and gradients are evaluated using overlaps computed on the IBM Belem quantum processor's simulator. In Fig.~\ref{fig:method1} we show the estimated value of the element of $W_{R/I}$ versus number of iterations of the classical optimization procedure, panels (a) and (b), and the values of the angles $\alpha_i$ which parametrize the trial eigenfunctions, panels (c) to (f). In the left panels, (a, c, e), we consider the real part of the observable $W$, i.e. $W_R$, while in the right panels (b, d, f), the imaginary part, $W_I$. In each of the two cases we show the results from $150$ runs of the protocol with initial angles for the two eigenfunctions $\alpha_1$ and $\alpha_2$ chosen uniformily between $-\pi$ and $\pi$. The line in each of the panels is obtained by the median value of the expectation value or angle between $50$ runs which end in the same vicinity. The colored background reflects the value taken by the middle $92\% $ of the corresponding realizations. In panels (a) and (b) we observe that while in the initial steps the angles cover the all range from $-\pi$ to $\pi$, and the expectation values take a very large range of possible values, within $20$ iterations the prediction of the matrix elements is very accurate, both for the real, panel (a), and imaginary part, panel (b).      
We note that the accuracy in the angles may not be as good as that on the matrix elements. This is one advantage of using this variational approach which is tailored to give the matrix elements directly. 

As a technical, but important, detail, for $W_R$ we had to fix the global phase of the trial eigenfunctions, otherwise the expectation value may show the wrong sign. For this reason when plotting the angles and evaluating the corresponding matrix elements, we mapped them between the angles $-\pi/2$ and $\pi/2$ so that the cosine of the angle would be positive. Such procedure is not needed for the imaginary elements of $W$, and for this reason the plotted range of angles in panels (d) and (f) is between $0$ and $2\pi$.     

\section{Analysis of errors for the single qubit case} 

\begin{figure}[htp] 
\centering
\includegraphics[width=\columnwidth]{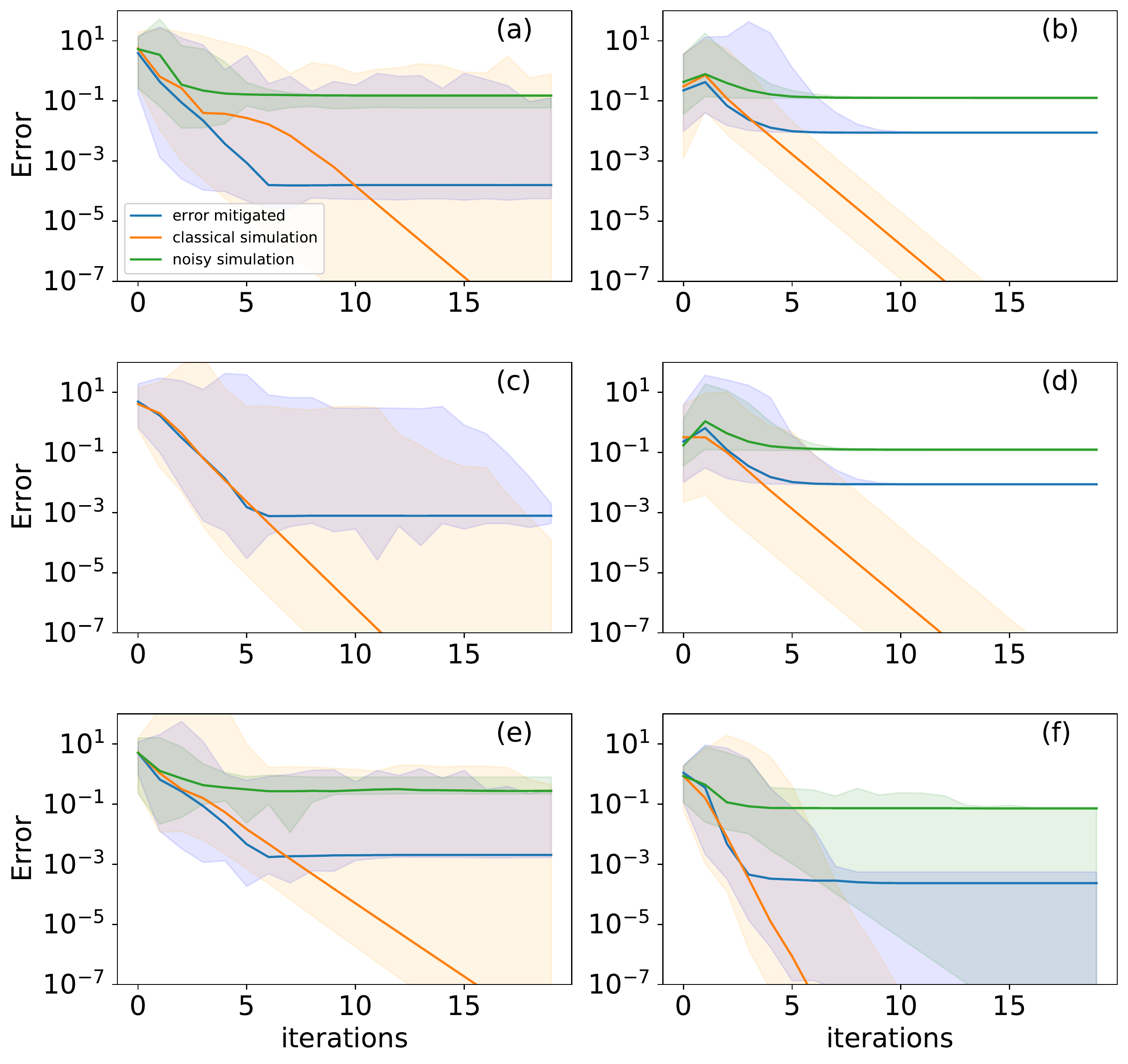}
\caption{Absolute value of the errors for each matrix element along 20 iterations, for classical, noisy and error mitigated simulations, along with the $92\%$ confidence level of the points. Both columns refer to experiments where the Lagrange multipliers $\ket{L_{1,\nu}}$ and $\ket{L_{2,\nu}}$ are computed exactly as discussed in Sec.~\ref{sec:algo}. Each panel (a-f) corresponds respectively to the matrix elements (in the energy eigenbasis) $2$, $2\im$, $5$, $-2\im$, $3$ and $0$. 
The orange continous line correspons to classical simulations, the blue dashed line to hybrid classical-quantum simulations with error mitigation, and the green dot-dashed line to hybrid classical-quantum simulations with no error mitigation. 
The lack of a green dot-dashed line in panel (c) implies that this element did not appear in our attempts hybrid classical-quantum simulations with no error mitigation.} 
\label{fig:method3}
\end{figure}   

For the single qubit case, both for the real and imaginary part we expect to obtain three different numbers from the variational approach. In Fig.\ref{fig:method3} we consider, in each panel, the error from each of these six possible values. More specifically we consider the real values 2, 5 and 3 in panels (a), (c) and (e) and the values from the imaginary part $2\im$, $-2\im$ and $0$ in panels (b), (d) and (f). In each panel we show the median of the $50$ runs approaching that value for three different cases: completely classical simulations (continuous orange lines), hybrid classical-quantum simulations without error mitigation (dashed green lines) and with error mitigation (dot-dashed blue lines). Also in this case the shadowing represent the $92\% $ confidence interval of the corresponding runs (i.e. between the $4-$th and the $96-$th percentile). We observe that only the fully classical approach is able to reach very small errors and continuously decrease as the iterations of the optimization routine increase. At the same time, also for the classical approach the process shows a non-negligible error bar. For the hybrid approach, instead, we observe that the error reaches a plateau after about $10$ iterations. This error is reduced when employing error-mitigation techniques. We thus associate this performance to an erroneous evaluation of the overlaps for the gradients. In some cases, as for the value $5$, the errors are so important that the hybrid classical-quantum algorithm is not able to converge to this solution when one does not implement error mitigation.

\subsection{Single qubit using an approximate evaluation of the Lagrange multipliers}

\begin{figure}[htp]
\includegraphics[width=\columnwidth]{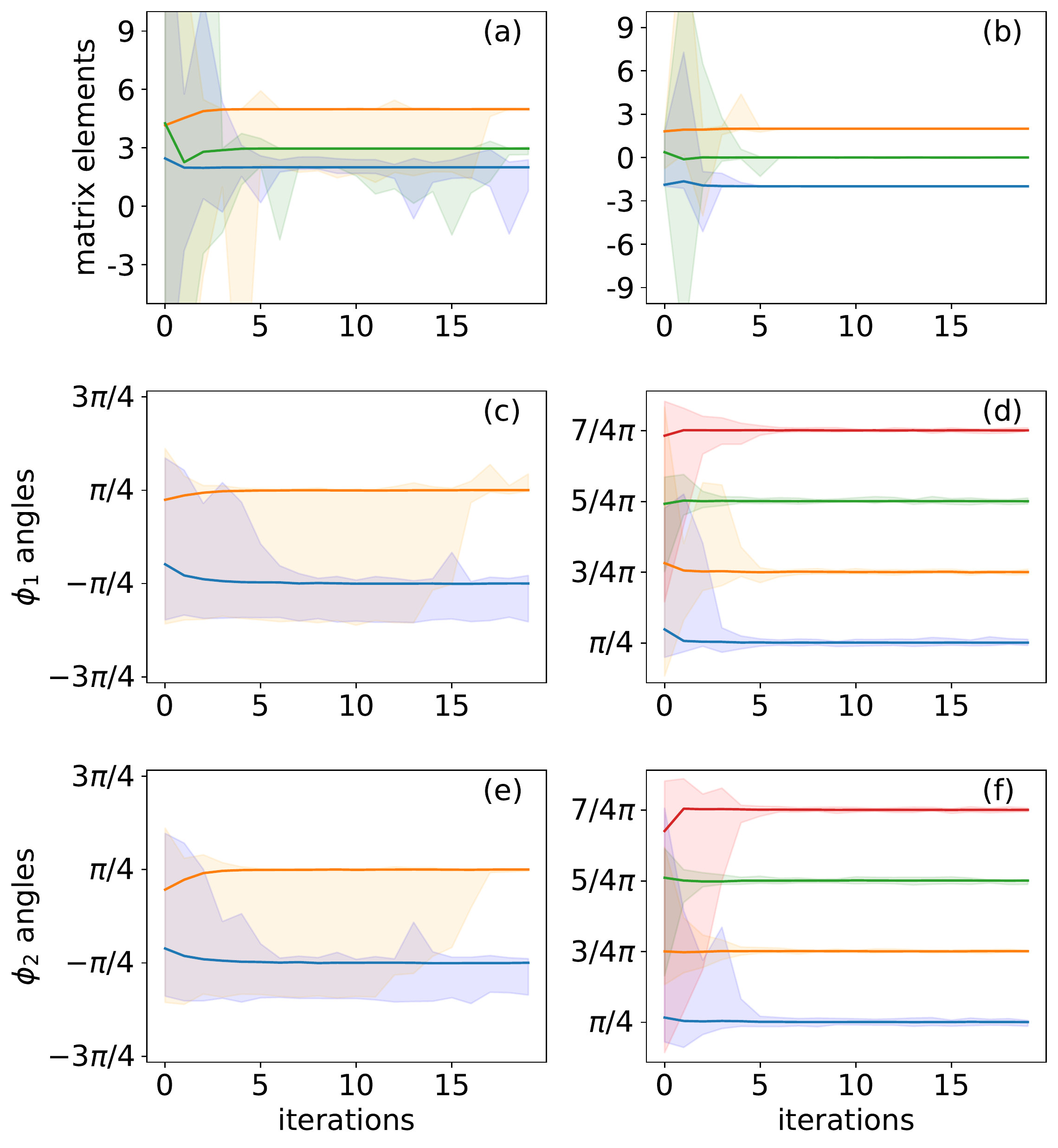}
\caption{Exactly the same description as Fig.~\ref{fig:method1} with the only difference that in the optimization procedure we have used approximated Lagrange multipliers from minimization of Eq.(\ref{eq:M_for_L_re}) or Eq.(\ref{eq:M_for_L_im}).}
\label{fig:method2}
\end{figure} 

Fig.~\ref{fig:method2} is completely analogous to Fig.~\ref{fig:method1}, and thus it shows results for real and imaginary elements of $W$ and the corresponding angles versus the number of iterations. However in this case we used the iterative algorithm from finding the minima of Eq.(\ref{eq:M_for_L_re}) and Eq.(\ref{eq:M_for_L_im}). Since in this case the Lagrange multipliers $|L_{i,\nu}\rangle$ are not exact, the optimization is not as accurate and hence while the medians of the expectation values and of the angles approach the exact values, the error bars, here also indicated by the $92\% $ confidence level of the runs ending close to one solution, are larger.

%%%%%%%%%%%%%%%%%%%%%%%%%%%%%%%% 
%%%%%%%%%%%%%%%%%%%%%%%%%%%%%%%% 

\section{Two qubits} 

\begin{figure}[htp]    
\centering
\includegraphics[width=\columnwidth]{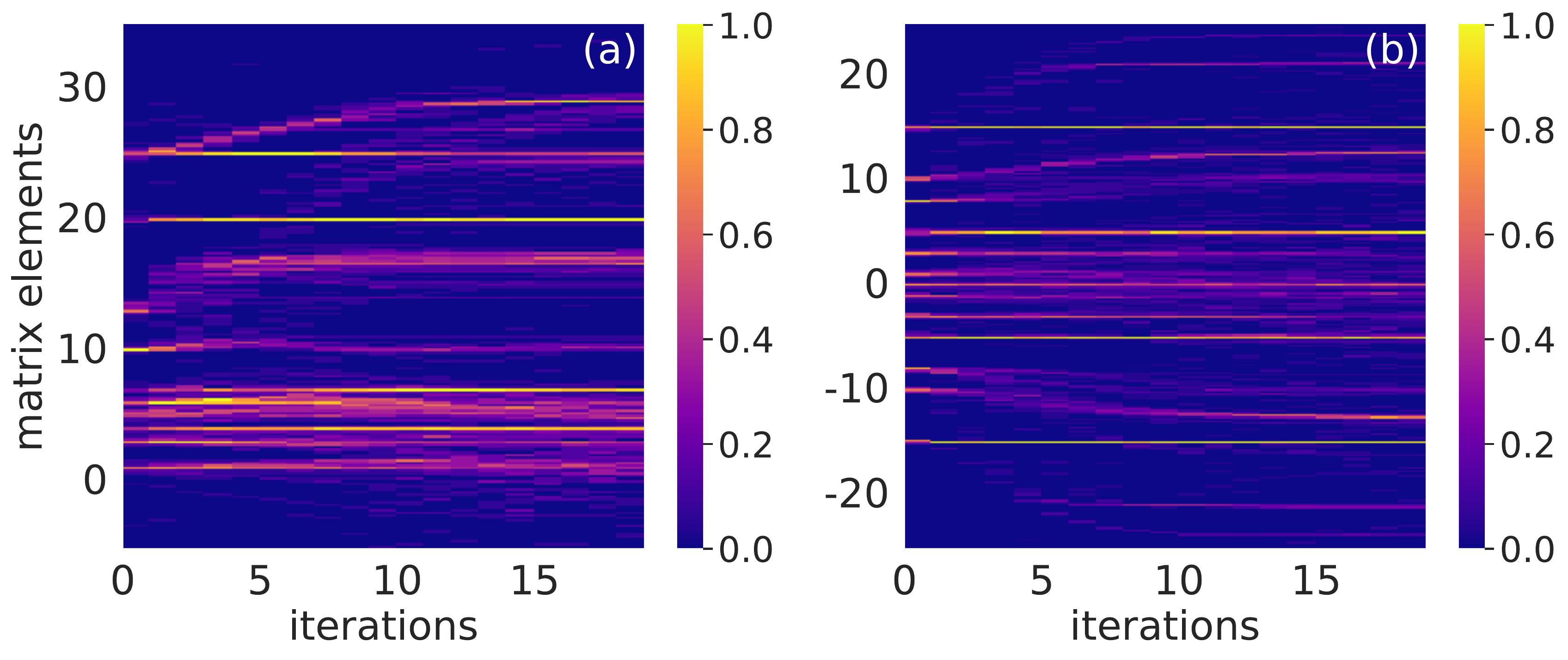}
\caption{Heatmaps for the two-qubit case and for the real (a) and imaginary (b) elements of $W_2$ in energy eigenbasis, through exact calculation of the Lagrange multipliers. Both subplots show the convergence to the matrix elements for 300 runs of 20 iterations each, initialized in intervals of radius 0.15 of the optimal angles. }
\label{fig:method4}
\end{figure}

To understand how this variational approach would perform on larger systems, we now consider a two-qubit case. We will use the Hamiltonian $H_2$ from Eq.(\ref{eq:two_qubits_ham}) and the observable $W_2$ which in the eigenbasis of the Hamiltonian has the values $W^D_2$ from Eq.(\ref{eq:Wd2}). The trial eigenfunctions are parametrized as shown in Eq.(\ref{eq: two_qubit_ansatz}). We consider $300$ different initializations and we show, in Fig.~\ref{fig:method4} a density plot of the resulting matrix elements vs the number of iterations. By density plot we mean that for each of the 20 iterations, we consider a vertical range and divide it into small bins of length 0.2 and count how many points fall into those intervals. The initial conditions are prepared such that near the exact angles for the solutions of the eigenfunctions in Eq.(\ref{eq: two_qubit_ansatz}) with an error $\pm 0.15$ from the exact $\alpha_i$, $\beta_i$ and $\gamma_i$. Furthermore, noises from the quantum machine have not been considered in these two-qubit calculations, and only statistical errors, i.e. shot-noise, are considered.     
From Fig.~\ref{fig:method4} we find that some matrix elements are much more stable than others. For instance the elements $20$, $25$ and $6$ converge in few iterations and they have a large probability of appearing. Other values, like $13$, are unstable and they do not appear unless one chooses initial conditions for the parameters of the trial eigenfunctions very close to the exact ones. This is true also for completely classical simulations, hinting at the fact that the landscape of this optimization problem is particularly difficult and better classical optimization routines should be used. What is possibly more striking is that there are also converged results to values which do not belong to $W_2$, as for instance the value $\approx 30$. From this we deduce that this method can give a good number of matrix elements, and depending on the amount of noise in the machine, they can be fairly accurate, but it may also not find some values and possibly return a few which are not correct. It would thus be important to cross-check these values with other methods.

\section{Conclusions} 

We have considered a variational approach for hybrid classical-quantum computation of the matrix elements of an observable $W$ in the energy eigenbasis. The diagonal elements of local observables typically relaxes to its long time asymptotic value. The fluctuations around this value as well as it relaxation times are however dictated by the off-diagonal elements \cite{beugeling2015off}.  Ths our approach has implications on  non-equilibrium statistical  studies. At the beginning, we have implemented the method with one qubit to learn some of its basic features, and then we proceed to two qubits to study how the method performs when the size of the systems studied increases. We have found that in general the method can perform well, meaning that it finds many of the matrix elements even when one initializes randomly the trial functions over a very broad range of parameters. The performance of the method are limited by the errors in the estimation of the gradients and overlaps from the quantum computer. These errors can be mitigated but they are still important, due to the fact that the variational function is not very stable close to some matrix elements. Furthermore, the variational function may also return some values that are not elements of the observable $W$. This occurs also in a fully classical implementation of the algorithm, indicating that improvements can be sought after with the landscape of the variational function and also with the classical optimizer. 
We highlight that one can use a much smaller basis, which represent states closer to those of interest, thus limiting the emergence of spurious incorrect results. In this case the optimization procedure will also be better controlled. 
Another way to improve the performance is to re-initialize the angles from the results after a certain number of iterations. %As we show in the Appendix, one can achieve quickly very good convergence \dpc{I am not sure we should show this}. 
Here we used normalized trial functions, but the method can be generalized, as we show in App.\ref{app:vp_nonnorma}, to non-normalized trial states. 
A relevant direction for future investigation is to find ways to stabilize the less stable matrix elements.

\begin{acknowledgments}
K.B.~acknowledges funding by U.S.~Department of Energy Award No.~DE-SC0019449, DoE ASCR Accelerated Research in Quantum Computing program (award No.~DE-SC0020312), DoE QSA, NSF QLCI (award No.~OMA-2120757), NSF PFCQC program, DoE ASCR Quantum Testbed Pathfinder program (award No.~DE-SC0019040), AFOSR, ARO MURI, AFOSR MURI, and DARPA SAVaNT ADVENT. KLC acknowledges support from the Ministry of Education and the National Research Foundation, Singapore.
D.P. acknowledges support from NRF-ISF grant NRF2020-NRF-ISF004-3528. 
\end{acknowledgments}

\bibliographystyle{apsrev4-1}
\bibliography{reference}

\appendix 
\section*{Appendix}

\section{Variational principle}\label{app:vp} 
The variational principle is best gleaned from examples.  Here, we restrict ourselves to the determination of off-diagonal matrix elements in quantum mechanics. In order to do so, we write the expression of the variational principle starting from the trial objective and the constraints, i.e. 
\begin{equation}
     F_v =\bra{\phi_{i,t}} W\ket{\phi_{j,t}}
\label{eq:1}
\end{equation}
and 
\begin{equation}
    B_{1}=(H-E_{i,j})\ket{\phi_{i,j}}=0 \text{ and } B_{1}^{\dagger}=\bra{\phi_{i,j}}(H-E_{i,j})=0.  
\label{eq:2}
\end{equation}
 \\
The final expression is obtained by promoting the constraints to the objective by multiplication by some Lagrange multipliers $L$ and $\lambda$. The goal is then to find expressions for $L$ and $\lambda$ either exactly or iteratively, and to check which ones are best suited for the variational principle. \\
 Concretely, the trial quantities of interest and the trial Lagrange multipliers are defined as 
 \begin{equation}
 \ket{\phi_{i/j,t}}=\ket{\phi_{i/j}}+\ket{\delta\phi_{i/j}},  
 \label{eq:3}
 \end{equation}
\begin{equation}
 \lambda_t=\lambda+\delta\lambda,     
 \label{eq:4}
 \end{equation} 
 and 
 \begin{equation}
 \ket{L_{i/j,t}}=\ket{L_{i/j}}+\ket{\delta L_{i/j}}.     
 \label{eq:5}
 \end{equation}
From this, we write the variational form of F, $F_v$, as 
\begin{align}
\begin{split}
     F_v &=\bra{\phi_{i,t}} W\ket{\phi_{j,t}} \\  &+\bra{L_{i,a,t}}(H-E_i)\ket{\phi_{i,t}} +\bra{\phi_{i,t}}(H-E_i)^{\dagger}\ket{L_{i,b,t}}  \\ & +\bra{L_{j,a,t}}(H-E_j)\ket{\phi_{j,t}} +\bra{\phi_{j,t}}(H-E_j)^{\dagger}\ket{L_{j,b,t}} \\\ &+\lambda\left[\bra{\phi_{i,t}} W\ket{\phi_{j,t}}\mp \bra{\phi_{j,t}} W\ket{\phi_{i,t}} \right],   
\label{eq:6}
\end{split}
\end{align}
and by replacing the trial quantities with equations \ref{eq:3}-\ref{eq:5}, one can get the error $\delta F_v$ as \\
$\delta F_v=F_v-\bra{\phi_{i,t}}W\ket{\phi_{j,t}}=(\bra{\phi_i}+\bra{\delta\phi_i})W(\ket{\phi_j}+\ket{\delta\phi_j})-\bra{\phi_i}W\ket{\phi_j}+(\bra{L_{i,a}}+\bra{\delta L_{i,a}})[(H-E_i)(\ket{\phi_{i}}+\ket{\delta \phi_i})]+[(\bra{\phi_i}+\bra{\delta \phi_i})(H-E_i)^{\dagger}](\ket{\delta L_{i,b}}+\ket{L_{i,b}})+(\bra{L_{j,a}}+\bra{\delta L_{j,a}})[(H-E_j)(\ket{\phi_{j}}+\ket{\delta \phi_j})]+[(\bra{\phi_j}+\bra{\delta \phi_j})(H-E_j)^{\dagger}](\ket{\delta L_{j,b}}+\ket{L_{j,b}})+
(\lambda+\delta \lambda)((\bra{\phi_i}+\bra{\delta\phi_i})W(\ket{\phi_j}+\ket{\delta\phi_j})\mp(\bra{\phi_j}+\bra{\delta\phi_j})W(\ket{\phi_i}+\ket{\delta\phi_i}))=0$
By using constraints \ref{eq:2} and \ref{eq:3},discarding terms of second order and putting equal to 0 the coefficients of $\ket{\delta \phi}$ and $\bra{\delta \phi}$, $\delta F_v$ vanishes for all allowed $\ket{\delta \phi}$ and $\bra{\delta \phi}$, and we get, for the real case
\begin{align}
\begin{split}
    &\bra{\delta\phi_i} \rightarrow (H-E_i)\ket{L_{i,b}}=
    -(\lambda+1)W\ket{\phi_j}, 
\label{eq:7}   
\end{split}
\end{align}
\begin{align}
\begin{split}
    & \bra{\delta\phi_j} \rightarrow (H-E_j)\ket{L_{j,b}}=\lambda W\ket{\phi_i},  
\label{eq:8}  
\end{split}
\end{align}
\begin{align}
\begin{split}
    &\ket{\delta\phi_i} \rightarrow \bra{L_{i,a}}(H-E_i)=
    \lambda \bra{\phi_j}W
\label{eq:9}   
\end{split}
\end{align} 
and 
\begin{align}
\begin{split}
    &\ket{\delta\phi_j} \rightarrow \bra{L_{j,a}}(H-E_j)=
    -(\lambda +1) \bra{\phi_i}W. 
\label{eq:10}   
\end{split}
\end{align}
Multiplying on the left of equation \ref{eq:7} by $\bra{\phi_i}$ and on the left of equation \ref{eq:8} by $\bra{\phi_j}$, and applying constraint \ref{eq:2}, we get $\lambda=-1/2$. $\ket{L}$ can be obtained by $\ket{L}=c_{1}\ket{\phi}$ and made unique by $\bra{\phi}\ket{L}=c_{2}=1$. We can see that for the real case $\ket{L_{i,a}}=\ket{L_{i,b}}$ and $\ket{L_{j,a}}=\ket{L_{j,b}}$. 

For pure imaginary matrix elements, the procedure is the same, and we get 
\begin{align}
\begin{split}
    &\bra{\delta\phi_i} \rightarrow (H-E_i)\ket{L_{i,b}}=
    -(\lambda+1)W\ket{\phi_j}, 
\label{eq:11}   
\end{split}
\end{align}
\begin{align}
\begin{split}
    & \bra{\delta\phi_j} \rightarrow (H-E_j)\ket{L_{j,b}}=-\lambda W\ket{\phi_i},  
\label{eq:12}  
\end{split}
\end{align}
\begin{align}
\begin{split}
    &\ket{\delta\phi_i} \rightarrow \bra{L_{i,a}}(H-E_i)=
    -\lambda \bra{\phi_j}W
\label{eq:13}   
\end{split}
\end{align} 
and 
\begin{align}
\begin{split}
    &\ket{\delta\phi_j} \rightarrow \bra{L_{j,a}}(H-E_j)=
    -(\lambda +1) \bra{\phi_i}W. 
\label{eq:14}   
\end{split}
\end{align}
Again, $\lambda=-1/2$ but this time $\ket{L_{i,a}}=-\ket{L_{i,b}}$ and $\ket{L_{j,a}}=-\ket{L_{j,b}}$. 

\section{Variational principle for $L_{i,\nu}$}\label{app:vpL}
The next step is to derive an expression, exact or approximated, for computing the Lagrange multiplier $\ket{L_{i,\nu}}$. This could in theory be done exactly by inverting $(H-E_{i})$ in \ref{eq:7}.
In practice, a near-singularity problem arises because the operator $H-E_1$, where $E_1$ is the ground state energy, has a zero eigenvalue and therefore cannot be inverted; this can be solved by replacing said operator by a shifted one that does not have a zero eigenvalue as discussed in the main text. In addition, in order to get an extremum principle that would give us an approximation to $\ket{L_{i,\nu}}$ without inverting any matrix, this operator must be positive definite. 

An operator that satisfies this condition is $H_{mod,i}-E_{1,t}=H-\frac{HP_{1,t}H}{E_{1,t}}-E_{1,t}$ where $P_{1,t}$ is the trial projection operator to the ground state $P_{1,t}=\ket{\phi_{1,t}}\bra{\phi_{1,t}}$ and $E_{1,t}$ is the trial ground state energy $E_{1,t}=\bra{\phi_{1,t}}H\ket{\phi_{1,t}}$. 
The idea is to find another variational principle for $\ket{L_{t}}$ (where we omit the indexes $i$ and $\nu$ to lighten the notation) that would replace Eq.~(\ref{eq:11}-\ref{eq:14}) , and that could then be used in the variational principle of Eq.~(\ref{eq:6}) to find better approximations of the trial functions. 
This variational principle for $\ket{L_{t}}$ is of the form 
\begin{equation}
    M(\ket{X_{tt}})=\bra{X_{tt}}A\ket{X_{tt}}-\bra{X_{tt}}\ket{q_t} -\bra{q_t}\ket{X_{tt}}
\label{eq:15}
\end{equation}
where A will be the shifted non-negative operator, $\ket{X_t}$ is the trial Lagrange multiplier $\ket{L_{tt}}$, $\ket{q_t}=A\ket{X_t}$ a known function and we write $\ket{X_{tt}}=\ket{X_t} +\ket{\delta X_t}$. By using this last equivalence, eq \ref{eq:15} becomes
\begin{equation}
    M(\ket{X_t} +\ket{\delta X_t})=M(\ket{X_{t}})+\bra{\delta X_t}A\ket{\delta X_t}
\label{eq:16}
\end{equation}
where the quadratic term on the far right is strictly convex since A is positive definite.
This expression has a minimum for $\ket{X_{tt}}=\ket{X_t}$, i.e. $\ket{L_{tt}}=\ket{L_{t}}$.\\
Setting $\ket{q_t}=A\ket{X_t}=(H_{mod,i}-E_{1,t})\ket{L_{t}}=[(\bra{\phi_{1t}}W\ket{\phi_{1t}})-E_{1t} c_{1t}]\ket{\phi_{1t}}-W\ket{\phi_{1t}}$, the variational principle for $\ket{L_{t}}$ becomes 
\begin{equation}
    M(L_{tt})=\bra{L_{tt}}(H_{mod,t}^{1}-E_{1t})\ket{L_{tt}}-\bra{L_{tt}}\ket{q_{t}}-\bra{q_{t}}\ket{L_{tt}}
\label{eq:17}
\end{equation}
and with $\ket{L_{tt}}=\ket{L_{t}}+\ket{\delta L_{t}}$,
\begin{equation}
    M(\ket{L_{t}}+\ket{\delta  L_{t}})=M_{11}(\ket{L_{t}})+\bra{\delta L_{t}}(H_{mod,t}^{1}-E_{1t})\ket{\delta L_{t}}
\label{eq:18}
\end{equation}
which has its minimum at $\ket{L_{tt}}=\ket{L_{t}}$.\\
For the real case, eq \ref{eq:17} becomes:
\begin{equation}
    M(L_{i/j})=\bra{L_{i/j}}(H_{mod,i/j}-E_{i/j})\ket{L_{i/j}}+\bra{\phi_{j/i}}W\ket{L_{i/j}}
\label{eq:19}
\end{equation}
while for the imaginary one, it is 
\begin{equation}
    M(L_{ib/jb})=\bra{L_{ib/jb}}(H_{mod,i/j}-E_{i/j})\ket{L_{ib/jb}}
\label{eq:19}
\end{equation}

\section{Variational principle for non normalized states}\label{app:vp_nonnorma}
The variational principle for non-normalized wave functions requires the introduction of the following additional constraints
\begin{equation}
    \bra{\phi_{i/j}}\ket{\phi_{i/j}}-1=0
\label{eq:20}
\end{equation}
and the respective Lagrange multipliers $\lambda_{i/j}$. 
The variational principle then becomes %\rebecca{can call $\lambda_1$ and $\lambda_2$: $\lambda_i$ and $\lambda_j$ ?}
\begin{align}
\begin{split}
     F_v &=\bra{\phi_{i,t}} W\ket{\phi_{j,t}} \\  &+\bra{L_{i,a,t}}(H-E_i)\ket{\phi_{i,t}} +\bra{\phi_{i,t}}(H-E_i)^{\dagger}\ket{L_{i,b,t}}  \\ & +\bra{L_{j,a,t}}(H-E_j)\ket{\phi_{j,t}} +\bra{\phi_{j,t}}(H-E_j)^{\dagger}\ket{L_{j,b,t}} \\\ & +\lambda_i (\bra{\phi_{i,t}}\ket{\phi_{i,t}}-1)+\lambda_j (\bra{\phi_{j,t}}\ket{\phi_{j,t}}-1)\\\ &
     +\lambda\left[\bra{\phi_{i,t}} W\ket{\phi_{j,t}}\mp \bra{\phi_{j,t}} W\ket{\phi_{i,t}} \right]  
\label{eq:22}
\end{split}
\end{align}
and we get the following equations for the real case 
\begin{align}
\begin{split}
    &\bra{\delta\phi_i}:(H-E_i)\ket{L_{i,b}}=-\lambda_i\ket{\phi_i}
    -(\lambda+1)W\ket{\phi_j}, 
\label{eq:23}   
\end{split}
\end{align}
\begin{align}
\begin{split}
    & \bra{\delta\phi_j}:(H-E_j)\ket{L_{j,b}}=-\lambda_j\ket{\phi_j}+\lambda W\ket{\phi_i},  
\label{eq:24}  
\end{split}
\end{align}
\begin{align}
\begin{split}
    &\ket{\delta\phi_i}: \bra{L_{i,a}}(H-E_i)=-\lambda_i\bra{\phi_i}+
    \lambda \bra{\phi_j}W, 
\label{eq:25}   
\end{split}
\end{align} 
and 
\begin{align}
\begin{split}
    &\ket{\delta\phi_j}: \bra{L_{j,a}}(H-E_j)=-\lambda_j\bra{\phi_j}
    -(\lambda +1) \bra{\phi_i}W
\label{eq:26}   
\end{split}
\end{align}
which give $\lambda=-1/2$ and $\lambda_i=\lambda_j=-(\bra{\phi_i}W\ket{\phi_j})/2$.
For the iterative approach, the function to optimize is 
\begin{align}
\begin{split}
    &M(L_{i/j})=\bra{L_{i/j}}(H_{mod,i/j}-E_{i/j})\ket{L_{i/j}}\\ &+2\lambda_{i/j} \bra{\phi_{i/j}}\ket{L_{i/j}}+\bra{\phi_{j/i}}W\ket{L_{i/j}}. 
\label{eq:27}
\end{split}
\end{align} 
For imaginary elements, we get 
\begin{align}
\begin{split}
    &\bra{\delta\phi_i}:(H-E_i)\ket{L_{i,b}}=-\lambda_i\ket{\phi_i}
    -(\lambda+1)W\ket{\phi_j}, 
\label{eq:28}   
\end{split}
\end{align}
\begin{align}
\begin{split}
    & \bra{\delta\phi_j}:(H-E_j)\ket{L_{j,b}}=-\lambda_j\ket{\phi_j}-\lambda W\ket{\phi_i},  
\label{eq:29}  
\end{split}
\end{align}
\begin{align}
\begin{split}
    &\ket{\delta\phi_i}: \bra{L_{i,a}}(H-E_i)=-\lambda_i\bra{\phi_i}
    -\lambda \bra{\phi_j}W, 
\label{eq:30}   
\end{split}
\end{align}
and 
\begin{align}
\begin{split}
    &\ket{\delta\phi_j}: \bra{L_{j,a}}(H-E_j)=-\lambda_j\bra{\phi_j}
    -(\lambda +1) \bra{\phi_i}W
\label{eq:31}   
\end{split}
\end{align}
with $\lambda=-1/2$ and $\lambda_i=\lambda_j=-(\bra{\phi_i}W\ket{\phi_j})/2$.\\
For the iterative approach for the imaginary elements, one needs to optimize 
\begin{align}
\begin{split}
    &M(L_{ib/jb})=\bra{L_{ib/jb}}(H_{mod,i/j}-E_{i/j})\ket{L_{ib/jb}}\\ &+2Re(\lambda_{i/j}) \bra{\phi_{i/j}}\ket{L_{ib/jb}}
\label{eq:32}
\end{split}
\end{align}

\end{document}